\documentclass[pdflatex,sn-nature]{sn-jnl}%

\usepackage{graphicx}%
\usepackage{multirow}%
\usepackage{amsmath,amssymb,amsfonts}%
\usepackage{amsthm}%
\usepackage{mathrsfs}%
\usepackage[title]{appendix}%
\usepackage{xcolor}%
\usepackage{textcomp}%
\usepackage{manyfoot}%
\usepackage{array}%
\usepackage{booktabs}%
\usepackage{algorithm}%
\usepackage{algorithmicx}%
\usepackage{algpseudocode}%
\usepackage{listings}%
\usepackage[inline]{enumitem}

\theoremstyle{thmstyleone}%

\theoremstyle{thmstyletwo}%

\theoremstyle{thmstylethree}%

\unnumbered%

\begin{document}

\title[Expertise capture]{Expertise capture in workplace AI}

\author*[1]{\fnm{Andreas} \sur{Ekelhart}}\email{andreas.ekelhart@univie.ac.at}

\author[2]{\fnm{Ahmad} \sur{Mohsin}}\email{a.mohsin@ecu.edu.au}

\author[2]{\fnm{Iqbal H.} \sur{Sarker}}\email{m.sarker@ecu.edu.au}

\author[2]{\fnm{Helge} \sur{Janicke}}\email{h.janicke@ecu.edu.au}

\affil*[1]{\orgdiv{Faculty of Computer Science}, \orgname{University of Vienna}, \orgaddress{\street{Kolingasse 16--19}, \city{Vienna}, \postcode{1090}, \country{Austria}}}

\affil[2]{\orgname{Edith Cowan University}, \orgaddress{\street{270 Joondalup Drive}, \city{Perth}, \state{WA}, \country{Australia}}}

\abstract{AI assistants are becoming part of how knowledge workers think and perform, accumulating the prompts, exemplars, memory and interaction histories that encode a worker's professional judgment. This infrastructure typically runs under employer or platform control. We name the resulting process expertise capture, the absorption of worker-applied knowledge and judgment into AI resources under organizational control. We develop a framework of contribution forms, technical residences and transformation families, and show that how far a contribution can be attributed, separated and revoked depends on where it resides and how it has been transformed, properties fixed by system design long before any dispute arises. When a worker departs, the organization retains a capability shaped by their judgment while the worker loses access to it, and the asymmetry compounds as contributions move from inspectable artifacts into derivatives, learned components and successor systems. We outline three measures, write-time provenance, extraction of separable contributions and an executable exit procedure, that make an agreed allocation technically performable, and make explicit what cannot be separated at all.}


\maketitle

\section{Introduction}\label{sec1}
Clark and Chalmers~\cite{clark1998extended} argued that a reliable, routinely endorsed external resource can function as part of a person's cognitive process rather than as a mere aid. AI assistants in the workplace increasingly meet these conditions~\cite{clark2025extending}. They accumulate persistent memory, curated prompts, exemplars, and interaction histories that encode practical judgment, while studies document productivity effects across writing, programming, and customer support~\cite{brynjolfsson2025generative}. This article addresses the issue that the applied expertise that makes such systems useful can be absorbed into technical resources that remain usable by the organization or platform after the worker has left.
In Clark and Chalmers' thought experiment, Otto owns the notebook that extends his memory. Workplace assistants usually run in an employer environment or platform tenant, under access controls and retention policies that the worker may not set~\cite{staufer2026agentindex}.
The central problem is not ordinary ownership of paid work product. It is architectural absorption: expertise paid for to be applied in work becomes persistent and difficult to separate from organizational resources.
Meta's program recording employees' screens and keystrokes to train task-performing agents has been widely discussed as workplace surveillance~\cite{cnbc2026meta}. But the issue runs deeper than monitoring, and it does not end if monitoring stops.

This process relates to conventional knowledge management and organizational
memory, through which organizations create, retain, and reuse knowledge beyond
the participation of individual workers~\cite{leoni2022knowledge}. Expertise capture addresses a narrower governance concern: although workers retain their embodied knowledge and professional judgment, externalized expressions of that expertise may remain operationally available through organizational AI resources after they leave.
Working effectively with an assistant means teaching it by supplying judgment, corrections, and examples. These persist in prompts, exemplars, workflow specifications, retrieval stores, memory, and logs, even without model training~\cite{lewis2020rag,park2023generativeagents}. The resulting resource is jointly produced~\cite{viljoen2021relational}: worker-applied expertise is entangled with organizational data, client obligations and employer-funded work. A fair settlement first requires systems that can distinguish transferable craft, organizational process, confidential context and technically inseparable residue.

We call this process \emph{expertise capture}: the absorption of worker-applied knowledge and professional judgment into AI resources controlled through organizational infrastructure. This work develops the concept and maps the forms expertise takes, the technical residences in which it persists, the transformations through which it moves, and the conditions under which applied expertise can be separated from organizational data and secrets.
This Perspective makes three contributions: (i) it distinguishes expertise capture from surveillance, ordinary ownership of work product, and model training by focusing on the persistence of applied judgment, (ii) it connects contribution forms, technical residences, and transformation families into a lifecycle account of how source coupling declines as lineage dependence rises, and (iii) it identifies provenance, extraction, and exit as design measures that make later legal, contractual, or organizational allocation technically implementable.

\section{The extended worker}\label{sec2}
The relationship between knowledge workers and AI assistants has moved beyond tool use. Across occupations, workers increasingly supervise and delegate: software developers specify, evaluate and steer coding agents, while deployed agentic systems are becoming capable of broader professional tasks~\cite{staufer2026agentindex}.
What matters for expertise capture is retention. A refined prompt encodes role-specific judgment, and an exemplar library preserves standards of quality. Agent memory carries decisions, preferences and situational context across sessions~\cite{park2023generativeagents,wang2024surveyagents}. For cognitive extension, the relevant conditions are reliability, routine availability and default endorsement~\cite{clark1998extended,clark2025extending}. Workplace assistants are beginning to satisfy these conditions without being treated literally as part of the worker's mind. It is enough that they become part of accumulated practice. This relation is reciprocal: recent work~\cite{vaccaro2024combinations} shows that the value of human-AI combinations varies by task.

More direct forms of augmentation, including brain-computer interfaces for cognitive enhancement~\cite{gordon2024bci}, may raise related questions about where organizational and personal cognition meet. The present argument concerns the nearer case: software assistants in organizational or platform tenants, where control over the infrastructure lies with the organization, not the worker.

Customer support gives the clearest empirical demonstration~\cite{brynjolfsson2025generative}. In a rollout across more than 5{,}000 support workers, an assistant trained on conversations of successful workers diffused tacit conversational patterns to the rest of the workforce. Productivity rose by 15\% on average, with the largest gains among novice and lower-skilled workers. The study shows skilled workers' know-how becoming an organizational resource that trained their colleagues.

Where AI-assisted performance becomes an organizational norm, workers may trade off withholding expertise against using tools that improve immediate performance. A correction or a curated example may affect a single interaction, but it can also be logged, converted into feedback data and used to improve a shared resource~\cite{ouyang2022training}. When workers expect their contributions to be reused against them, they begin to withhold knowledge~\cite{cullen2026labor}. Organizational research defines knowledge hiding and withholding as established workplace behaviours~\cite{connelly2012knowledgehiding}, and recent work links AI-related insecurity and automation threat to those behaviours~\cite{liu2025robotknocks}. For assistant-mediated work, however, withholding may mean abandoning the tools through which the work is organized. The question is technical as well as institutional: where does expertise reside, how does it move, and what separation is implementable?

\section{A framework for expertise capture}\label{sec3}
Not every system that stores organizational information captures expertise. Three elements must come together: a worker contributes judgment, corrections, examples, procedures or patterns of choice; the contribution becomes operationally persistent as memory, retrieval content, prompt material, log data or a learned component; and the resulting resource is held under organizational infrastructure or governance after the worker stops participating.

\begin{figure}[t]
\centering
\includegraphics[width=0.98\textwidth]{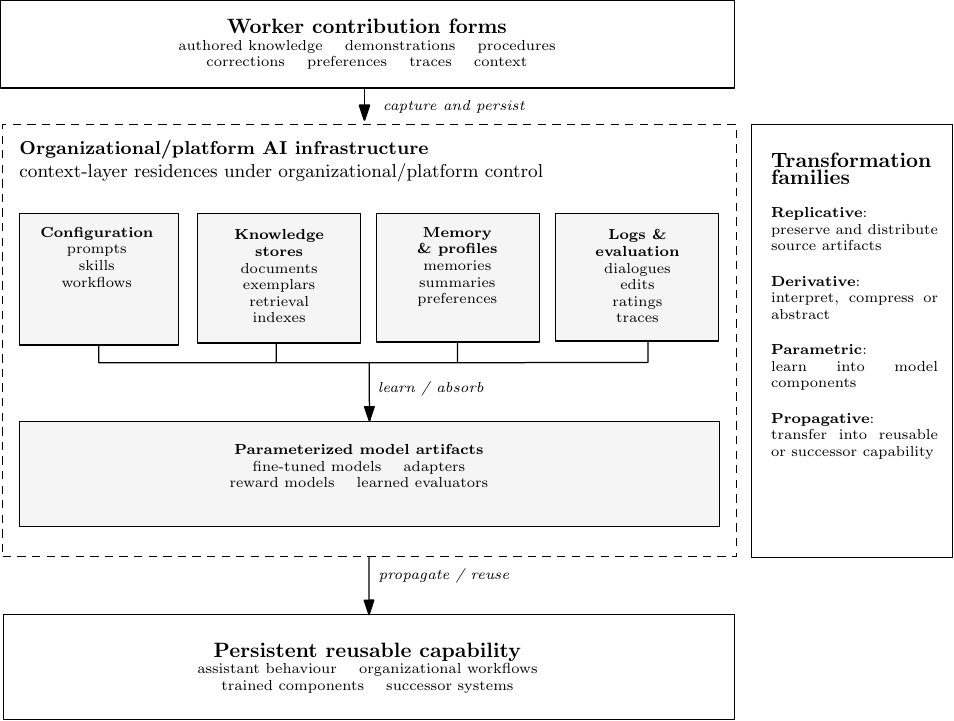}
\caption{Lifecycle view of workplace expertise capture. Worker contributions are captured in context-layer residences, may be absorbed into parameter-layer artifacts and can propagate into reusable or successor capabilities. The transformation families describe how the relation between the contribution and the worker changes across this lifecycle.}\label{fig:architecture}
\end{figure}

Figure~\ref{fig:architecture} presents the framework as a lifecycle: contributions from ordinary assistant-mediated work persist in context-layer residences, may be learned into parameter-layer artifacts and can become reusable capability under organizational or platform control. A correction, for example, can become a log entry, then a preference dataset, a tuned adapter and eventually a successor assistant. At each step the same contribution becomes harder to attribute, separate or reverse~\cite{adhikesaven2026models,xu2026governance}. The framework sits inside a wider socio-technical system involving workers, deploying organizations, providers and, in some cases, clients or data subjects. Its purpose is to show why technical form determines what can later be traced, separated, removed or exported.

\subsection{Contribution forms}
A worker may supply authored knowledge, demonstrations, exemplars, workflows, corrections, rankings, preferences, interaction traces and context. These forms support different later claims. For example, an authored prompt states an inspectable rule, a demonstration shows how a task is performed, and a selected exemplar records a quality standard. Repeated accept--reject decisions reveal judgment through use, and traces may expose procedures never separately articulated. Expertise capture therefore begins before model training, since routine work generates persistent resources by being recorded, summarized, indexed or reused.

\subsection{Technical residences}
The first distinction is between the context layer and the parameter layer. The context layer contains artifacts supplied to a model at run time: prompts, persona files, skills, workflow specifications and templates~\cite{wang2024surveyagents}. Knowledge stores hold documents, exemplar libraries, vector indexes built from embeddings~\cite{reimers2019sentence}, retrieval corpora~\cite{lewis2020rag}, knowledge graphs and summaries. Memory covers episodic, semantic and procedural memories in agent systems~\cite{park2023generativeagents}, inferred preferences~\cite{kirk2024personalizing} and session summaries~\cite{wang2024surveyagents}. Logs and evaluation data comprise conversation histories, tool traces, edits, accepted or rejected outputs, demonstrations, evaluation sets and preference datasets for human-feedback training~\cite{ouyang2022training}. These residences are often inspectable and organized as discrete items, but governance requires knowing who contributed each item, for what purpose and under what confidentiality. Dataset-provenance and responsible-dataset work documents the same gap~\cite{longpre2024datasetaudit}.

The parameter layer contains learned artifacts. Workplace expertise may enter through continued or domain-adaptive pretraining~\cite{gururangan2020dontstop}, supervised fine-tuning and preference optimization~\cite{ouyang2022training}, reward-model training~\cite{ouyang2022training}, adapter training~\cite{hu2022lora}, learned evaluators or routers. Once absorbed into parameters, contributions generally cease to be separable artifacts. A worker may be known to have contributed to a corpus, but the causal effect on model behaviour is difficult to estimate~\cite{koh2017influence} and harder to remove~\cite{liu2025rethinking}. Modular components such as adapters can be deleted as units, but the contributions of individual workers cannot be separated within them~\cite{hu2022lora}.

\subsection{Transformation families}
Residences describe where captured expertise persists, while transformations describe how its relation to the worker changes. The families below are not an exhaustive ontology of AI engineering operations but a grouping of transformations by their governance effect on source coupling, provenance, attribution, separability, and revocability.

\paragraph*{Replicative.}
Replicative transformations preserve and distribute source artifacts through ingestion, copying, chunking, indexing, caching, format conversion and migration, operations that are routine in retrieval-augmented systems~\cite{lewis2020rag,reimers2019sentence}. The problem is proliferation: the contribution remains recognizable, but copies, chunks, indexes and caches may no longer be enumerable.

\paragraph*{Derivative.}
Derivative transformations create artifacts by interpreting, compressing, generalizing or inferring from sources.
Embeddings~\cite{reimers2019sentence}, retrieval summaries~\cite{lewis2020rag}, memory extraction and reflective agent traces~\cite{park2023generativeagents}, profile inference~\cite{staab2024privacyinference}, knowledge-graph extraction, generated workflows~\cite{wang2024surveyagents} and synthetic examples fall into this category.
The problem is semantic fusion~\cite{staab2024privacyinference,shilov2026mosaic}: a derivative may preserve transferable judgment while mixing it with client context, organizational process, or the system's own inferences.

\paragraph*{Parametric.}
Parametric transformations absorb patterns into learned components through pretraining~\cite{gururangan2020dontstop}, fine-tuning~\cite{ouyang2022training}, preference optimization~\cite{ouyang2022training}, reward-model training~\cite{ouyang2022training}, adapters~\cite{hu2022lora} and related methods.
The contribution no longer persists as a discrete object. Attribution requires external training records and influence estimates~\cite{koh2017influence}, and deletion requires approximate unlearning methods~\cite{liu2025rethinking}. These methods do not provide general guarantees of contributor-level attribution or removal.

\paragraph*{Propagative.}
Propagative transformations reproduce an already-derived capability in another resource, through distillation~\cite{hinton2015distilling}, model merging, synthetic-output training, workflow transfer between agents~\cite{wang2024surveyagents} or successor dependencies in model pipelines~\cite{adhikesaven2026models}. The problem is lineage distance: a later model, dataset or agent may carry a worker-derived capability without containing the original material, so governance constraints need to survive downstream transformation~\cite{xu2026governance}.

Together these form a gradient from source preserved, to source interpreted, absorbed and reproduced (Table~\ref{tab:transformations}). Across that gradient, source coupling declines, downstream persistence increases and governance depends increasingly on external lineage records~\cite{moreau2013provdm} rather than on the artifact itself.

\begin{table}[t]
\caption{Transformation families and their governance properties. For each family, the table summarizes the relation between derivative and source, typical operations, and the resulting conditions for provenance, attribution, separability and revocability of a worker's contribution. These are technical tendencies rather than fixed legal entitlements.}\label{tab:transformations}
\begingroup
\newcommand{\rotlabel}[1]{\makebox[0.052\textwidth][c]{\rotatebox[origin=c]{90}{#1}}}
\setlength{\tabcolsep}{2pt}
\renewcommand{\arraystretch}{1.08}
\scriptsize
\begin{tabular}{@{}>{\centering\arraybackslash}m{0.052\textwidth}>{\raggedright\arraybackslash}p{0.226\textwidth}>{\raggedright\arraybackslash}p{0.226\textwidth}>{\raggedright\arraybackslash}p{0.226\textwidth}>{\raggedright\arraybackslash}p{0.226\textwidth}@{}}
\toprule
 & Replicative & Derivative & Parametric & Propagative \\
\midrule
\rotlabel{Source relation} & Most coupled: source preserved and distributed & Source interpreted, compressed or abstracted & Source absorbed into learned components & Least coupled: capability reproduced in successor resources \\
\midrule
\rotlabel{Operations} & Ingestion, copying, chunking, indexing, caching & Embeddings, summaries, memory extraction, profile inference, generated workflows & Continued pretraining, fine-tuning, preference optimization, reward-model and adapter training & Distillation, model merging, synthetic-data training, successor deployment \\
\midrule
\rotlabel{Provenance} & Direct if identifiers and copy records are preserved & Lineage-dependent across one or more sources & External training and corpus records required & End-to-end lineage required across systems \\
\midrule
\rotlabel{Attribution} & Usually high for known instances & Conditional and often many-to-one & Approximate and contested at the contributor level & Very low without successor lineage \\
\midrule
\rotlabel{Separability} & High for known instances; copies and caches complicate removal & Artifact separability may remain; semantic separability decreases where sources are fused & Low at contributor level; modular components such as adapters can be deleted only as units & Low where successors no longer contain the original source \\
\midrule
\rotlabel{Revocability} & Feasible for known instances; incomplete without copy records & Derivative-dependent and difficult once fused into profiles or summaries & Limited without retraining or validated unlearning & Downstream-dependent; source removal may not affect successors \\
\botrule
\end{tabular}
\endgroup
\end{table}

\subsection{Worker expertise to persistent capability}
\label{sec:worked-example}
\begin{figure}[t]
\centering
\includegraphics[width=\textwidth]{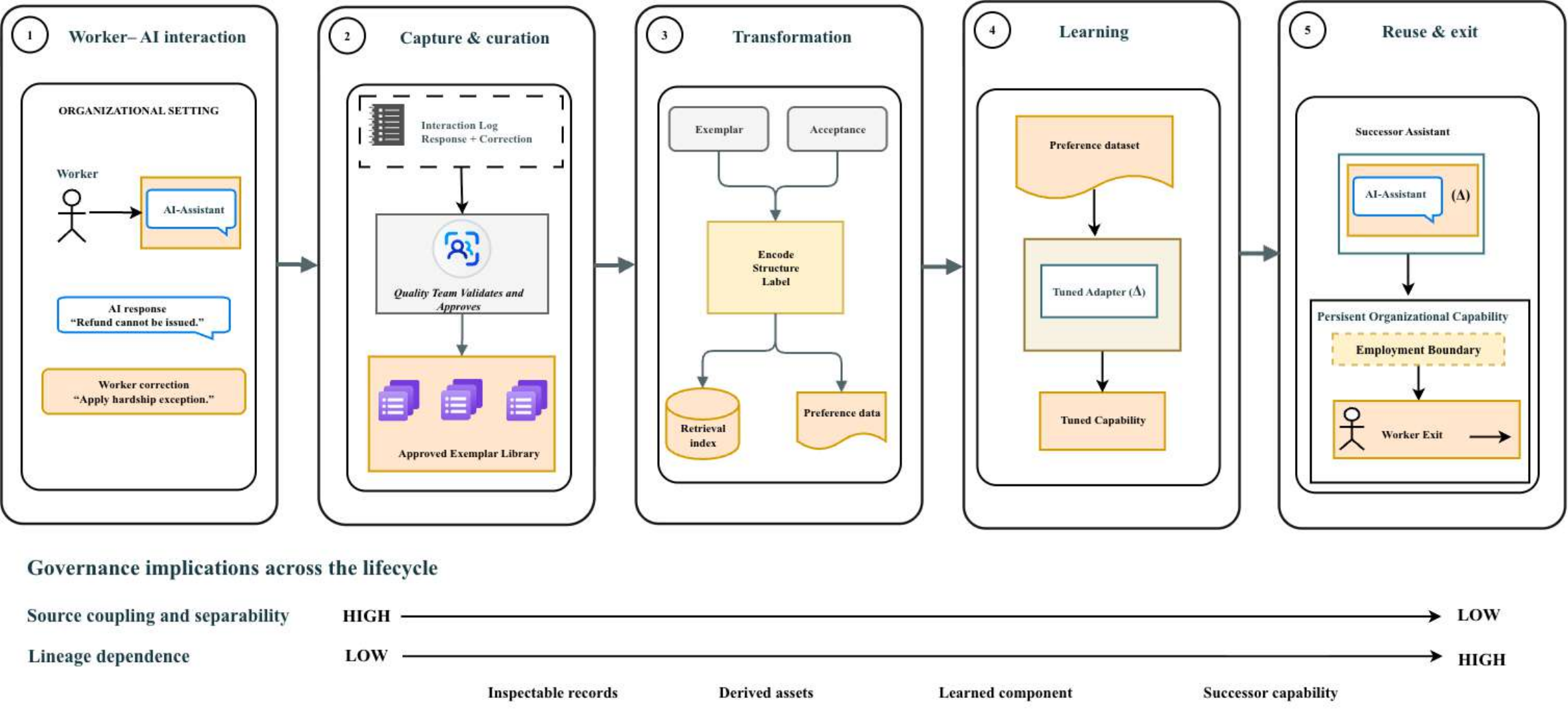}
\caption{Worker-applied expertise transformed into a persistent organizational AI capability. The five stages instantiate the transformation families of Table~\ref{tab:transformations}: capture and curation are replicative, encoding and preference construction derivative, adapter tuning parametric, and successor integration propagative.}
\label{fig:worked-example}
\end{figure}

A running customer-support example shows how a worker's intervention can become a persistent organizational AI capability (Figure~\ref{fig:worked-example}). The worker replaces an unsuitable AI response with an approved hardship exception. The initial response, revised answer, and attribution metadata are stored in an interaction log, reviewed by a quality team, and converted into an approved exemplar. Research on sentence embeddings~\cite{reimers2019sentence} and retrieval-augmented generation~\cite{lewis2020rag} supports embedding and indexing this exemplar for later retrieval. Research on human-feedback training~\cite{ouyang2022training} similarly shows how the initial and revised responses can form a preference pair. This preference data may support low-rank adapter tuning~\cite{hu2022lora}, allowing the resulting adapter to be integrated into a successor assistant.

This sequence instantiates the four transformation families of Table~\ref{tab:transformations}. Record retention is replicative. Curation, embedding and preference construction are derivative. Adapter tuning is parametric, and successor integration is propagative. If the worker leaves, they retain their embodied expertise but lose access to the assistant-based resources, while the organization continues using the resulting capability. Research on model dependencies~\cite{adhikesaven2026models} and governance across model derivations~\cite{xu2026governance} supports the concern that such contributions become harder to trace and separate as they move through these transformations.

\section{Expertise capture and asymmetric control}\label{sec4}
The consequences of capture are most visible when employment ends, but may arise earlier as captured capabilities alter how work is evaluated, allocated, and performed.
Captured expertise is co-produced: workers develop judgment through organizational resources, colleagues, clients and institutional context, so no simple claim of individual ownership survives.
The organization's interests are equally real. It funds the infrastructure, bears client and confidentiality obligations, and has its own stake in process knowledge and pooled capability.
Neither side's claim is enforceable over an undifferentiated resource, which is why separability matters to both.
Employment law, confidentiality, trade secrets, data protection, worker data rights, and AI-specific regulation each raise distinct questions~\cite{abraha2025workerdata}. Publicity and personality interests arise where AI systems create digital replicas~\cite{liang2025clones}. Copyright questions about AI training remain unsettled~\cite{uscopyrightoffice2025training}. This article does not settle these legal questions.
The prior technical question is whether a contribution can be found, separated, removed or exported once an allocation is chosen.

In many workplace deployments, organizations control the environment in which the technical residences of captured expertise persist after departure~\cite{staufer2026agentindex,abraha2025workerdata}. Not everything in these residences should be portable, since they also hold organizational process, client information and collective work. The asymmetry is that the worker may lose access to externalized judgment, while the organization continues to use a capability shaped by it. Portability is also a question of professional continuity. A worker's judgment and methods have always left with the worker, carried in memory and in personal notes.
Externalized into an assistant under organizational control, the same methods
no longer leave by default.

Exit requirements depend on the captured capability. A person-specific configuration, clone, avatar or digital replica that imitates an identifiable worker's style, voice, image, preferences or decision patterns raises identity, privacy and representation concerns. May an organization continue to present, consult or trade on a recognizable simulation of someone no longer there~\cite{liang2025clones}? A pooled capability built from many workers' contributions raises collective concerns instead, about substitution, bargaining power and distribution of gains across an occupational group~\cite{viljoen2021relational,cullen2026labor}.
The boundary between the two cases is continuous, but the safeguards differ.

Separability makes these safeguards implementable. The question is whether worker-applied expertise can be distinguished from organizational process, confidential context and other contributors' work. Explicit configuration artifacts such as prompts, persona files, skills and templates state rules, standards and procedures as reviewable text. They can often be redacted so transferable craft remains while client names, figures and internal procedures stay behind. Retrieval stores are item structured because documents are ingested, chunked, embedded and indexed as retrievable units~\cite{reimers2019sentence,lewis2020rag}. Filtering is plausible only when ingestion records contributor, source, purpose and confidentiality metadata, as data-lineage models and ML-provenance systems do~\cite{moreau2013provdm,schlegel2025mlprovenance}. Without that metadata, public references and confidential client deliverables can sit side by side in a vector store with no reliable basis for deciding what may leave.

Personal data stores (PDS) suggest one design route. In consumer settings, PDS architectures give individuals a controlled repository for personal data and mediate service access~\cite{fallatah2023pds}. Solid-style pods and task-driven data capsules show more specific mechanisms for consent-mediated access and revocation~\cite{mansour2016solid}. A workplace analogue would be a worker-controlled or jointly governed expertise store. Prompts, exemplars, portable lessons and preference records could be stored in a separable residence, while project facts and client material remain in organizational stores. This would not solve allocation, and workplace data rights remain constrained by employment power asymmetries~\cite{abraha2025workerdata}, but it would make separation possible.

The harder cases are memory, logs and learned components. Agent memory entries and interaction logs arise as byproducts of work and can fuse judgment with the client names, figures and context that occasioned it~\cite{park2023generativeagents}. Redaction may destroy the lesson or preserve the lesson while leaking context, a trade-off familiar from text anonymization, sanitization and privacy inference~\cite{pilan2022tab,staab2024privacyinference}. Where worker data trains or tunes a model, the contribution exists only as a diffuse change in weights. There is no artifact to extract, and current LLM unlearning is not a realistic substitute for contributor-level deletion in ordinary enterprise systems. Full retraining is costly and requires retained clean corpora, while approximate unlearning can damage retained capabilities and is hard to verify~\cite{liu2025rethinking}. Even when a capability is suppressed, it may be bypassed or audited as residual memory~\cite{shilov2026mosaic}.

Transformations intensify the asymmetry because contributions do not remain in their initial residence.
A correction can become a log entry, the log a preference dataset, the dataset a tuned model and the model a distilled successor. Protections attached to the first artifact do not follow automatically. A ninety-day log-retention rule says little about a model tuned on the logs in week two~\cite{xu2026governance}. Governance must bind to contributions as they move through transformations and successors. Dataset and model documentation already make provenance a governance norm~\cite{gebru2021datasheets}. ML provenance and post-training lineage work show why the chain must record who supplied each item and where it came from, into which residence, for what purpose, under what confidentiality, through which transformations and into which successors~\cite{schlegel2025mlprovenance,adhikesaven2026models}. The chain is metadata rather than content, written as each step runs and maintained outside the derivative artifact. The next section turns this into three design measures: write-time provenance, extraction of separable contributions, and an executable exit procedure.

\section{Provenance, extraction and exit as ways forward}\label{sec5}
\begin{figure}[h]
\centering
\includegraphics[width=0.98\textwidth]{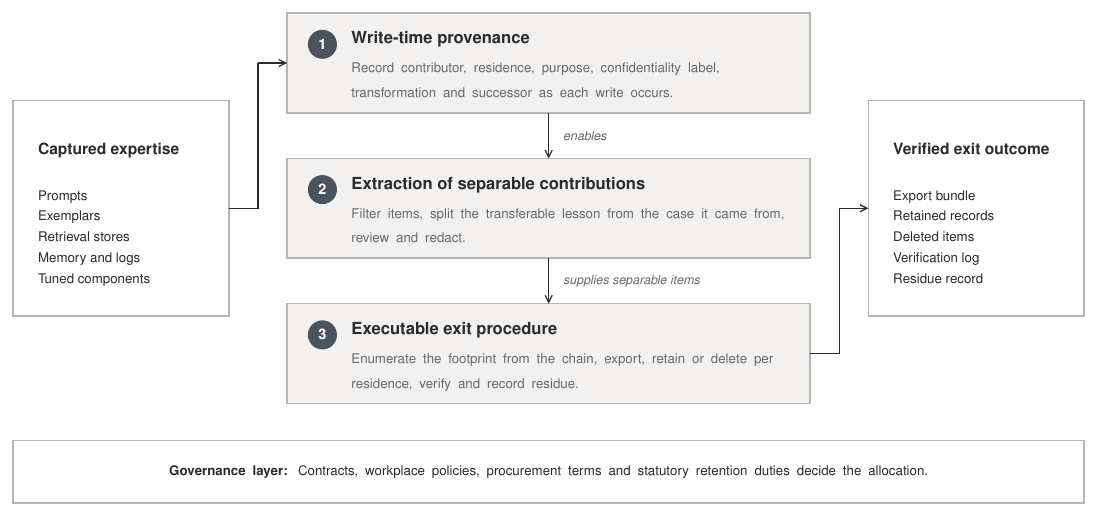}
\caption{Operational view of the three design measures. Write-time provenance makes the captured expertise footprint enumerable, extraction of separable contributions distinguishes transferable craft from protected context, and an executable exit procedure applies and verifies the agreed allocation.}\label{fig:exitprocedure}
\end{figure}

The properties that decide practical control are fixed when systems are built, not when disputes arise. Provenance, extraction and exit are minimum design requirements for whatever allocation a contract, regulator, collective agreement or court later adopts. They are not a complete governance architecture, and they do not decide who should own, use or remove a captured resource. Instead, they make such decisions executable. Figure~\ref{fig:exitprocedure} summarizes the sequence.

The first measure is write-time provenance. Entry and transformation events are discrete engineering steps, and linking them to contributors as they run is far more feasible than reconstructing them after derivatives, caches and learned components have accumulated.
The minimum record contains contributor, residence, source, purpose, confidentiality label, transformation event and successor link. That record helps protect company secrets while allowing transferable expertise to be identified later.

Where a platform or model provider is separate from the employer, privacy-enhancing techniques such as federated learning~\cite{mcmahan2017communication}, differential privacy~\cite{abadi2016deep} and secure or encrypted computation~\cite{kaissis2021endtoend} can reduce what a supplier learns from organizational data or employee traces. They complement provenance rather than replace it: exposure is limited during computation, but these techniques do not decide who may use a captured capability, whether a worker may export a memory, or how a derivative should be treated at exit.

The second measure is extraction of separable contributions. A departing worker's transferable prompts, exemplars, skills and memory fragments can be exported only if the system supports item-level filtering, review and redaction. Extraction is narrower than portability: not every worker-touched resource leaves, but transferable craft can be distinguished from organizational process, client context and shared work.
For agent memory, the split must happen when entries are written. One approach is to split each memory into linked parts, a transferable lesson, such as a rule of thumb or preference, and the specific case it came from. The lesson can then be exported while client names and figures remain behind. Whether this can be done reliably remains open.

Because memory schemas, retrieval interfaces and transformation logs are product internals, extraction often depends on platform or model vendors~\cite{staufer2026agentindex}. That creates an employer-side problem as well as a worker-side one: organizations may need portability and audit rights to avoid lock-in to a vendor that cannot enumerate or divide captured expertise.

Provenance records who supplied an artifact and what was derived from it, not what the artifact contains. Confidentiality boundaries often run within artifacts: one prompt can state a transferable rule and a client's name in one sentence. Splitting content is therefore separate from tracing it. Classification methods, including named-entity and sensitive-data detection~\cite{lison2021anonymisation}, data-loss-prevention filters and model-assisted review, can help identify candidate sensitive material. Human review must still decide whether statements are transferable craft, organizational process or client material. Redaction removes specifics, while generalization or synthetic replacement substitutes artificial values. Both carry fidelity costs and residual-leakage risks~\cite{pilan2022tab,staab2024privacyinference}. For weights, membership and extraction probes can test whether material is reproducible from a model~\cite{shokri2017membership,carlini2021extracting}, while influence estimates~\cite{koh2017influence} can support coarse attribution, though none yields guarantees~\cite{liu2025rethinking}. These are supporting techniques, not solutions. Different claims require different levels of evidence.

The third measure is an executable exit procedure. A departing worker's footprint is enumerable only through the provenance chain. Without it, neither party can list what is in dispute. How the footprint is divided is legal and contractual. Technically, the procedure needs two capabilities. First, the provenance chain must produce a complete inventory. Second, each residence must support the chosen action: remove or export store items, copy or redact configuration files, delete adapters where appropriate, or record irremovable residue in weights. The result must be verifiable: bundle completeness is checked against the provenance chain, and cleanliness by re-scanning exported material and testing residual identification~\cite{pilan2022tab,staab2024privacyinference}.

These measures exist in current platforms only in fragments. A deployable architecture would need four components. Ingestion and logging pipelines record contributors and confidentiality. Transformation records link corpora, tuned models and successors to sources. Extraction interfaces filter stores and route authored texts through review. Exit procedures enumerate a contributor's footprint across derivatives. These components can be built with known techniques, but become substantially harder to implement once capture has occurred.

Organizations have reason to build this early rather than after a dispute. Without such records they cannot enumerate what a departing worker contributed, show that client material did not leave with an export, or divide captured expertise held in vendor internals. Workers who believe that the assistant they must use weakens their bargaining position may also withhold knowledge~\cite{cullen2026labor} or shift professional extension into personal assistants outside company control.

A remaining open question is whether accumulated judgment can be separated from the specific contexts that produced it.
No technical mechanism settles how footprints are divided or under whose custody the records sit. Contracts, procurement terms, workplace policies and collective procedures must specify what the architecture implements. Employee data rights add another layer.
The GDPR right to data portability (Article 20), for example, is the closest existing instrument to extraction but covers only personal data provided by the data subject, leaving derived and inferred artifacts uncovered~\cite{gdpr2016,wp29portability}.
The right to erasure (Article 17) is subject to exceptions where processing is required by law or legal claims, and employment, pension, tax, audit and regulated-record obligations may require retention even when export or deletion is requested~\cite{gdpr2016,abraha2025workerdata}.
The EU AI Act~\cite{aiact_2024} adds record-keeping duties, but only for high-risk AI systems and at system rather than contributor level, so ordinary workplace assistants and worker-level provenance fall outside it.
Expertise capture therefore opens a multidisciplinary agenda for security and privacy engineering, law, ethics, management and organizational psychology: technical separability makes fair governance possible, but does not determine what fairness requires.

\section{Conclusion}\label{sec6}
Knowledge workers increasingly think and perform through AI assistants, and the infrastructure of that extension, prompts, exemplars, skills, retrieval stores, memory and logs, is where their expertise accumulates. We have proposed expertise capture as a framework of contribution forms, technical residences and transformation families, and shown that practical control, attribution, separability and revocability must be addressed at design time. The consequences materialize when an employee leaves: the organization retains a working capability, while the worker loses access to externalized judgment. They compound as contributions move from context-layer artifacts into derivatives, learned components and successor systems~\cite{adhikesaven2026models,xu2026governance}. Write-time provenance, extraction of separable contributions and an executable exit procedure make agreed arrangements technically implementable.

A workable governance agenda begins with technical architecture: preserve provenance, separate transferable craft from confidential context where possible, and make exit a procedure rather than a dispute reconstructed after the fact.

\backmatter

\section*{Acknowledgements}

A.E. acknowledges support by SBA Research (SBA-K1 NGC) a COMET Center within the COMET – Competence Centers for Excellent Technologies Programme and funded by BMIMI, BMWET, and the federal state of Vienna. The COMET Programme is managed by FFG.

\section*{Use of generative AI}

The authors used generative AI tools, including ChatGPT, Codex, and Claude to support manuscript drafting, structural revision, language editing, reference checking and figure preparation. The authors reviewed, edited and verified the outputs and take responsibility for the final content.

\bibliography{sn-bibliography}

\end{document}